\documentclass[openacc]{rsproca_new}

\DeclareMathOperator{\arctanh}{arctanh}

\usepackage{comment}
\usepackage{units}

\begin{document}

\title{Plasmonic modes in cylindrical nanoparticles and dimers}

\author{
Charles A.\ Downing$^{1,2}$ and Guillaume Weick$^{3}$}

\address{$^{1}$Departamento de F\'isica de la Materia Condensada, CSIC-Universidad de Zaragoza, Zaragoza E-50009, Spain\\
$^{2}$Department of Physics and Astronomy, University of Exeter, Exeter EX4 4QL, United Kingdom\\
$^{3}$Universit\'{e} de Strasbourg, CNRS, Institut de Physique et Chimie des Mat\'eriaux de Strasbourg,  UMR 7504, F-67000 Strasbourg, France}

\subject{solid state physics, nanotechnology, optics}


\corres{Guillaume Weick\\
\email{guillaume.weick@ipcms.unistra.fr}}

\begin{abstract}

We present analytical expressions for the resonance frequencies of the plasmonic modes hosted in a cylindrical nanoparticle within the quasistatic approximation. Our theoretical model gives us access to both the longitudinally and transversally polarized dipolar modes for a metallic cylinder with an arbitrary aspect ratio, which allows us to capture the physics of both plasmonic nanodisks and nanowires. We also calculate quantum mechanical corrections to these resonance frequencies due to the spill-out effect, which is of relevance for cylinders with nanometric dimensions. We go on to consider the coupling of localized surface plasmons in a dimer of cylindrical nanoparticles, which leads to collective plasmonic excitations. We extend our theoretical formalism to construct an analytical model of the dimer, describing the evolution with the inter-nanoparticle separation of the resultant bright and dark collective modes. We comment on the renormalization of the coupled mode frequencies due to the spill-out effect, and discuss some methods of experimental detection. 

\end{abstract}


\begin{fmtext}
\section{Introduction}
\label{Sec:intro}
The optical properties of small metal clusters have been studied throughout the 20th century~\cite{Kreibig1995}, in a field which is now referred to as plasmonics~\cite{Maier2007}. Modern nanoplasmonics aims to confine and control light at the nanoscale, in an amalgamation of photonics and \phantom{electronics}
\end{fmtext}


\maketitle

\noindent electronics~\cite{Ozbay2006}. It is envisaged that applications will arise in areas from data storage and microscopy to light generation and biophotonics~\cite{Barnes2003, Gramotnev2010, Stockman2011}. 
In the last few years, 
the subfield of quantum plasmonics 
has branched away, whereby quantum mechanical phenomena play a crucial role~\cite{Tame2013}.

An intensively studied quasiparticle in plasmonics is the localized surface plasmon (LSP), a collective oscillation of conduction band electrons, which arises when a metallic nanoparticle (NP) is irradiated by light~\cite{Maier2007} or hot electrons \cite{Abajo2010}. Exploring how the resonance frequency of the plasmon changes depending on the geometry of its hosting NP is a fundamental task of the field~\cite{Kerker1969, Hulst1981, Bohren1983, Mishchenko2000, Mishchenko2002, Mayergoyz2013}. Recently, a number of groundbreaking experiments~\cite{Schmidt2012, Juve2013, Schmidt2014, Krug2014, Wang2017, Movsesyan2018, Schmidt2018} have probed the plasmonic response of metallic cylinders, and in particular the limiting cases of nanodisks and nanowires. Inspired by these experiments, in this work we derive simple, analytical expressions for the dipolar plasmon resonances within the quasistatic limit (valid when the dimensions of the NP is smaller than the wavelength associated with the LSP resonance frequency) in both the longitudinal and transverse polarizations (that is, along the cylindrical axis and perpendicular to it). 
Our model is based upon a calculation of the change in Hartree energy of the NP due to the collective displacement of the valence electrons. We assume that the electrons in the nanostructure form a body of approximately uniform density, which allows us 
to employ continuum mechanics and 
set up a simple equation of motion~\cite{Bertsch1994}. 
Importantly, our analytic theory is valid for any aspect ratio of the cylinder, and as such is of relevance for a wide range of experiments. 
Our work therefore complements previous theoretical studies of plasmonic cylinders, which have either employed the nanowire 
approximation~\cite{Lukyanchuk2006, Abrashuly2019, Gangaraj2020}, or have required numerics~\cite{Schroter2001, Mayergoyz2005, Sburlan2006, Amendola2010, Tserkezis2012, Napoles2015}.

In our model, the inevitable quantum corrections which arise at the nanoscale are addressed by accounting for the so-called spill-out effect~\cite{Brack1993}. In this quantum size effect, the resonance frequency is modified due to a proportion of electrons spilling outside of the small metallic NP, thus lowering the average electronic density inside the NP. This effect arises due to the ground-state many-body wave function, which determines the electronic density, having tails which leak outside of the sharp boundary of the NP surface, so that a non-negligible number of electrons reside outside of the cluster. The spill-out effect has been studied historically in relation to spherical NPs~\cite{Bertsch1994}, and more recently has been investigated for plasmons in ultra-sharp groove arrays \cite{Skjolstrup2017}.

Coulomb interactions between LSPs housed in different NPs can give rise to collective plasmons spread out over the combined nanostructure~\cite{Prodan2003, Jain2010}. The study of collective plasmons in NP arrays, including architectures built from cylindrical NPs~\cite{Sahoo2009, Guillot2010, Zoric2011, Gillibert2017, Kawachiya2018}, has led to a wealth of diverse physics, from plasmonic waveguides~\cite{Maier2003, Brandstetter2016} to light harvesters~\cite{Cole2006, Aubry2010} to analogues of a topological insulator~\cite{Poddubny2014, Downing2017,DowningLamb2018,Pocock2018}. 
In this work, we are concerned with the simplest example of a coupled system, the NP dimer~\cite{Nordlander2004, Reinhard2005, Jain2007, Brandstetter2015}, which constitutes the building block of more complex metastructures, and where insight into the nature of coupled plasmons can be achieved.

A series of experiments on nanoplasmonic dimers in the near-field coupling regime have revealed both bright and dark plasmonic modes, where the dipole moments are oriented in-phase or out-of-phase, respectively~\cite{Chu2009, Koh2009, Barrow2014}. In order to account analytically for such collective plasmonic effects, we adapt our aforementioned theory to the case of a dimer of cylindrical metallic NPs. We derive simple expressions for the bright and dark mode resonance frequencies of the system as a function of the interparticle separation, which allows for a clear description of how the plasmonic coupling scales with distance. Our results supplement theories of cylindrical dimers in the literature, which predominately involve assumptions about the aspect ratio of the cylinder, or requires time-consuming numerical computations~\cite{Kottmann2001, Hoflich2009, Vorobev2010, Babicheva2012, Toscano2012, Schnitzer2016b, Bereza2019}. We also comment on the spill-out effect in the dimer, and suggest some methods for the experimental detection of our predicted effects.

This paper is organized as follows: in \S\ref{sec:1_cyl}, we calculate the dipolar resonances of a single cylindrical NP and discuss their respective decay rates. We find the modifications to the resonance frequencies due to the spill-out effect in \S\ref{Sec:single_cylinder_spill}. The theory is extended to describe collective effects in a dimer of cylindrical NPs in \S\ref{Sec:dimer}. Finally, we draw some conclusions in \S\ref{Sec:conc}.

\section{Plasmonic modes in a single cylindrical nanoparticle}
\label{sec:1_cyl}

\begin{figure*}[tb]
 \includegraphics[width=1.0\linewidth]{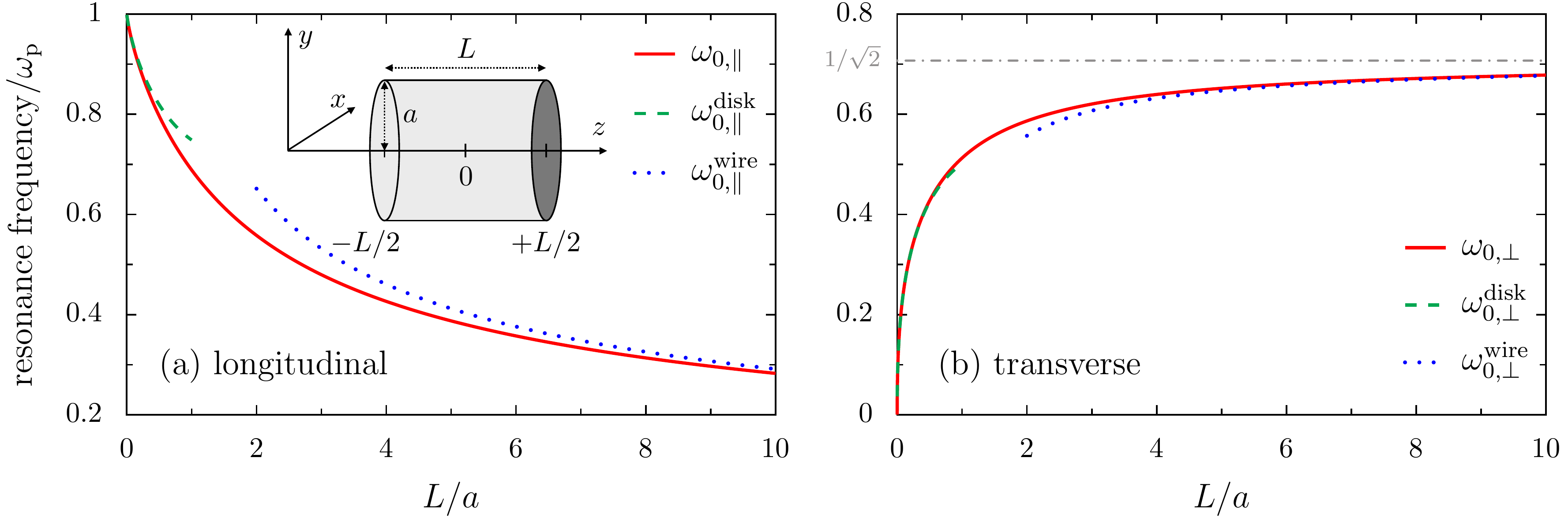}
 \caption{Resonance frequencies $\omega_{0,\parallel}$ and $\omega_{0,\perp}$ (solid red lines), in units of the plasma frequency $\omega_\mathrm{p}$, as a function of the aspect ratio $L/a$ for both the (a) longitudinal [cf.\ \eqref{eq:dimer_spectrum_1}] and (b) transverse [cf.\ \eqref{eq:Trans_modes}] modes. Dashed green lines: the disk limit approximations, from \eqref{eq:disk1} and \eqref{eq:Trans_limits_disk} for panels (a) and (b), respectively. Dotted blue lines: the wire limit approximations, from \eqref{eq:wire1} and \eqref{eq:Trans_limits_wire} for panels (a) and (b), respectively. Horizontal dash-dotted line in panel (b): the asymptotic result $ \omega_{\mathrm{p}}/\sqrt{2}$, for $L/a\to\infty$. 
 Inset: Sketch of a cylindrical metallic nanoparticle of radius $a$ and length $L$.} 
 \label{fig:frequencies_cylinder}
\end{figure*}

We consider a cylindrical NP of radius $a$ and length $L$, containing $N_\mathrm{e}$ valence electrons with charge $-e<0$ and mass $m_\mathrm{e}$ (see the inset in figure \ref{fig:frequencies_cylinder}). We start by neglecting the electronic spill-out effect, and assume that the density $n(\mathbf{r})$ of valence electrons is uniform (with density $n_0$) inside the cylinder, and vanishing outside, i.e.,
\begin{equation}
\label{eq:Long_Hartree}
 n (\mathbf{r}) = n_0\, \Theta \left(a-r\right) \Theta \left(\frac{L}{2} + z\right) \Theta \left(\frac{L}{2}-z\right), 
\end{equation}
where $(r, \theta, z)$ are the usual cylindrical coordinates, and where $\Theta(x)$ is the Heaviside step function.

Our strategy to obtain the frequencies of the plasmonic normal modes along the longitudinal ($\hat z$, $\alpha=\parallel$) and transverse ($\hat r$, $\alpha=\perp$) directions\footnote{Here and in what follows, hats designate unit vectors.} closely follows the one presented, e.g., in reference~\cite{Bertsch1994} for a spherical NP, which yields for the LSP resonance frequency the well-known Mie result $\omega_\mathrm{p}/\sqrt{3}$, with $\omega_\mathrm{p}$ the plasma frequency.\footnote{Note that a similar phenomenological approach has been successfully applied 
by the authors of reference \cite{Yin2009} to spin-dependent dipole excitations, and excellent agreement was obtained against time-dependent density functional theory numerical calculations.} 
We first impose  a rigid shift $\mathbf{u}_\alpha$ of the electron distribution, which gives rise to the displaced density  $n(\mathbf{r} - \mathbf{u}_\alpha)$.  Assuming that $u_\alpha=|\mathbf{u}_\alpha|$ is small with respect to the dimensions of the cylinder, we have $n(\mathbf{r} - \mathbf{u}_\alpha) \simeq n (\mathbf{r}) + \delta n_\alpha (\mathbf{r})$, with
\begin{equation}
\label{eq:Long_Hartree_delta}
\delta n_\alpha (\mathbf{r}) = - \mathbf{u}_\alpha \cdot \nabla n (\mathbf{r}).
\end{equation}
We then consider the resulting change in the Hartree energy (in cgs units)
\begin{equation}
\label{eq:Hartree_alpha}
 \delta {E}_\alpha = \frac{e^2}{2} \int \mathrm{d}^3\mathbf{r} \int\mathrm{d}^3\mathbf{r'}\, \frac{\delta n_\alpha (\mathbf{r}) \delta n_\alpha (\mathbf{r'})}{|\mathbf{r}-\mathbf{r'}|},
\end{equation}
with respect to the equilibrium situation. This quantity gives access to the restoring force 
\begin{equation}
\label{eq:restoring}
F_\alpha=-\frac{\partial}{\partial {u}_\alpha}\left(\delta E_\alpha\right)=-k_\alpha u_\alpha
\end{equation}
 and to the resulting spring constant $k_\alpha$. The latter quantity then provides an expression for the normal mode frequency 
\begin{equation}
\label{eq:normal_mode}
\omega_{0,\alpha}=\sqrt{\frac{k_\alpha}{M_\mathrm{e}}}, 
\end{equation}
where $M_\mathrm{e}=N_\mathrm{e}m_\mathrm{e}$ corresponds to the total electronic mass.

Let us now consider the longitudinal ($\alpha=\parallel$) and transverse ($\alpha=\perp$) polarizations each in turn, which arise from different electronic distribution displacements $\mathbf{u}_\alpha$.

\subsection{Longitudinal mode}
\label{Sec:single_cylinder_nospillout_long}

We assume the longitudinal displacement $\mathbf{u}_{\parallel} = u\, \hat{z}$, such that the change in density \eqref{eq:Long_Hartree_delta} is
\begin{equation}
\label{eq:Cylinder_delta}
 \delta n_{\parallel} (\mathbf{r}) = un_0\,  \Theta \left(a-r\right)
 \left[ \delta \left(z-\frac{L}{2}\right) - \delta \left(z+\frac{L}{2}\right)  \right],
\end{equation}
where $\delta(x)$ is the Dirac delta function. Equation \eqref{eq:Cylinder_delta} corresponds to a charge imbalance that is located at the two disks  of radius $a$ closing the cylinder at $z=\pm L/2$ (cf.\ the inset in figure \ref{fig:frequencies_cylinder}).
In order to evaluate the modification of the Hartree energy \eqref{eq:Hartree_alpha} due to the above density change, 
we shall exploit the Laplace expansion of the Newtonian kernel \cite{Jackson1998}
\begin{equation}
\label{eq:Cylinder_Laplace}
\frac{1}{|\mathbf{r}-\mathbf{r'}|}= \frac{2}{\pi} \sum_{m=-\infty}^{+\infty} \int_0^{\infty} \mathrm{d} k\, \mathrm{e}^{\mathrm{i} m (\theta-\theta')} \cos \left( k [z-z'] \right)  
 I_m \left( k r_{<}\right) K_m \left( k r_{>}\right).
\end{equation}
Here, $I_m (x)$ and $K_m (x)$ are modified Bessel functions of the first and second kinds, respectively, while $r_{<} = \mathrm{min} (r, r')$ and $r_{>} = \mathrm{max} (r, r')$. 
Upon inserting \eqref{eq:Cylinder_delta} and 
\eqref{eq:Cylinder_Laplace} into \eqref{eq:Hartree_alpha}, we arrive at a seven-dimensional integral. After carrying out the straightforward angular and Cartesian integrals, and using the following result for the double radial integral,
\begin{equation}
\label{eq:Long_result}
 \int_0^a\mathrm{d} r\, r \int_0^a\mathrm{d} r'\,   r' I_0 \left( k r_{<}\right)  K_0 \left( k r_{>}\right)  
  = \frac{a^2}{2k^2}  \left[ 1 - 2 I_1 (k a) K_1 (k a)  \right],
\end{equation}
we find 
\begin{equation}
\label{eq:Long_Hartree_2}
 \delta E_\parallel = 8 \pi (e n_0 u)^2 a^3 \int_0^{\infty}  \frac{\mathrm{d}x}{x^2}\, \sin^2 \left( \frac{L}{2 a} x \right) \left[ 1 - 2 I_1 (x) K_1 (x) \right], 
\end{equation}
which is harmonic in the displacement $u$.
Evaluating the first term in the above integral using $\int_0^{\infty}\mathrm{d}t\, \sin^2{(t)}/t^2  = \pi / 2$, and integrating the second term employing special functions, we find
\begin{equation}
\label{eq:Long_Elliptic}
 \delta E_{\parallel} = 4 \pi \left( e u n_0 \right)^2 a^3 \left[ \frac{\pi L}{2a} + \frac{4}{3} - g\left(\frac{L}{a}\right) \right].
\end{equation}
In the expression above, 
the function $g(x)$ is defined as
\begin{equation}
\label{eq:g_function}
 g(x) = \frac{x}{6} \left[ \left( x^2 + 4 \right) K\left( -\frac{4}{x^2} \right) - \left( x^2 - 4 \right) E\left( -\frac{4}{x^2} \right) \right],
\end{equation}
where
\begin{equation}
\label{eq:Def_elliptic}
 K(x) = \int_0^1 \frac{\mathrm{d} t}{\sqrt{(1-t^2)(1-xt^2)}}, \qquad\qquad  E(x) = \int_0^1 \mathrm{d} t\, \sqrt{\frac{1-xt^2}{1-t^2}}
\end{equation}
are the complete elliptic integrals of the first and second kinds, respectively. 
The monotonically increasing function \eqref{eq:g_function} has the following asymptotic expansions for small and large arguments:
\begin{subequations}
\label{eq:g_function_2}
\begin{align}
 g(x) &\simeq \frac{4}{3} + \left( 6 \ln{2} - 1 - 2 \ln{x} \right) \frac{x^2}{4} + \mathcal{O}(x^4), \quad x \ll 1, \label{eq:g_function_2a} \\
 g(x) &\simeq \frac{\pi}{2} \left( x + \frac{1}{2 x} - \frac{1}{4 x^3} \right) + \mathcal{O}(x^{-5}),  \quad x \gg 1. \label{eq:g_function_2b}
 \end{align}
\end{subequations} 
The result \eqref{eq:Long_Elliptic}, together with \eqref{eq:restoring} and \eqref{eq:normal_mode}, then yields the following analytic expression for the resonance frequency of the dipolar longitudinal mode of the cylinder:
\begin{equation}
\label{eq:dimer_spectrum_1}
\omega_{0,\parallel} = \omega_\mathrm{p} \sqrt{ 1 + \frac{2a}{\pi L} \left[ \frac{4}{3} - g \left( \frac{L}{a} \right) \right]}.
\end{equation}
Here the plasma frequency of the considered metal is
$\omega_\mathrm{p} =({4 \pi n_0 e^2}/{m_\mathrm{e}})^{1/2}$,
with the electron density $n_0=N_\mathrm{e}/\pi a^2L$ for the examined cylinder.

We plot in figure \ref{fig:frequencies_cylinder}(a) the longitudinal resonance frequency \eqref{eq:dimer_spectrum_1} as a function of the aspect ratio $L/a$ of the cylinder as the solid red line. As one can see from the figure, 
$\omega_{0,\parallel}$ is a monotonically decreasing function of the parameter $L/a$, with the limiting values 
$\lim_{L/a\to0}\{\omega_{0,\parallel}\}=\omega_\mathrm{p}$ and $\lim_{L/a\to\infty}\{\omega_{0,\parallel}\}=0$, which coincide with the well-known asymptotic results for a spheroidal NP \cite{Bohren1983}. 
Physically, the longitudinal mode softens when the aspect ratio of the cylinder increases, since the ratio of uncompensated charges to the compensated ones (by the ionic background) decreases with increasing $L/a$. We 
note that this trend has been confirmed experimentally \cite{Juve2013}.

We now consider the two limiting cases of \eqref{eq:dimer_spectrum_1}, namely when the cylinder can be treated as 
a nanodisk ($L/a\ll1$) or a nanowire ($L/a \gg 1$), and where insightful expressions can be obtained. Let us first 
examine the disk limit. Using the expansion \eqref{eq:g_function_2a}, 
\eqref{eq:dimer_spectrum_1} becomes
\begin{equation}
\label{eq:disk1}
\omega_{0,\parallel}^{\mathrm{disk}} \simeq \omega_\mathrm{p} \left\{ 1 - \frac{L}{4 \pi a} \left[6 \ln{2} - 1 - 2 \ln{\left( \frac{L}{a} \right)} \right] \right\},\qquad L/a\ll1.
\end{equation}
Clearly, this expression tends linearly towards the plasma frequency $\omega_\mathrm{p}$ in the extreme pancake limit ($L \to 0$), see the dashed green line in figure \ref{fig:frequencies_cylinder}(a). In the opposite limit of a wire, 
we obtain with~\eqref{eq:g_function_2b} 
\begin{equation}
\label{eq:wire1}
\omega_{0,\parallel}^{\mathrm{wire}} \simeq \omega_\mathrm{p} \sqrt{ \frac{8 a}{3 \pi L} },\qquad L/a\gg1,
\end{equation}
which is plotted as a blue dotted line in figure \ref{fig:frequencies_cylinder}(a), showcasing the inverse square root decay to zero frequency.

\subsection{Transverse mode}
\label{Sec:single_cylinder_nospillout_trans}

In order to have access to the eigenfrequency of the transverse dipolar plasmonic mode, here we assume the arbitrary small displacement $\mathbf{u}_{\perp} = u\, \hat{x}$ (see the inset in figure \ref{fig:frequencies_cylinder}), such that the change in electron density \eqref{eq:Long_Hartree_delta} is
\begin{equation}
\label{eq:Cylinder_delta_trans}
 \delta n_{\perp} (\mathbf{r}) = un_0  \cos{\theta}\, \delta \left(r-a\right) \Theta \left(\frac{L}{2}+z\right) \Theta \left(\frac{L}{2}-z\right).
\end{equation}
Completing an analogous calculation as to that for the preceding case of the longitudinally polarized mode [cf.\ 
\S\S2\ref{Sec:single_cylinder_nospillout_long}] leads to the following equation for the change in the Hartree 
energy~\eqref{eq:Hartree_alpha}, 
\begin{equation}
\label{eq:Trans_Hartree}
 \delta E_{\perp} =  8 \pi (e n_0 u)^2 a^3 \int_0^{\infty}  \frac{\mathrm{d}x}{x^2}\, \sin^2 \left( \frac{L}{2 a} x \right)  I_1 (x) K_1 (x).
\end{equation}
Evaluating the above integral then yields
\begin{equation}
\label{eq:Trans_Hartree_2}
 \delta E_{\perp} = 2 \pi \left( e u n_0 \right)^2 a^3 \left[ g\left(\frac{L}{a}\right) - \frac{4}{3}  \right],
\end{equation}
where $g(x)$ is defined in \eqref{eq:g_function}. 
We thus obtain an analytic expression for the resonance frequency of the transverse dipolar plasmonic mode, using \eqref{eq:restoring} and \eqref{eq:normal_mode} with \eqref{eq:Trans_Hartree_2}, as
\begin{equation}
\label{eq:Trans_modes}
\omega_{0,\perp} = \omega_\mathrm{p} \sqrt{\frac{a}{\pi L} \left[ g \left( \frac{L}{a} \right) -\frac{4}{3} \right] }.
\end{equation}

We plot the transverse resonance frequency \eqref{eq:Trans_modes} in figure \ref{fig:frequencies_cylinder}(b) as the
 solid red line, as a function of the aspect ratio $L/a$. As is evident from the figure, 
$\omega_{0,\perp}$ is a monotonically increasing function of the parameter $L/a$, bounded by the two limits
$\lim_{L/a\to0}\{\omega_{0,\perp}\}=0$ and $\lim_{L/a\to\infty}\{\omega_{0,\perp}\}=\omega_{\mathrm{p}}/\sqrt{2}$ (the latter limit is denoted by the horizontal dash-dotted line in the figure). 
As is the case for the longitudinal plasmonic mode, such asymptotic limits are the same for a spheroidal NP \cite{Bohren1983}.
Contrary to the longitudinal mode shown in figure \ref{fig:frequencies_cylinder}(a), the transverse mode gets harder when the aspect ratio of the cylinder increases, since the ratio of uncompensated charges that sit on the longitudinal surface of the cylinder to the compensated ones increases with increasing $L/a$.

In figure \ref{fig:frequencies_cylinder}(b), the limiting cases of a nanowire ($L/a\ll 1$, dashed green line) and nanodisk ($L/a\gg1$, dotted blue line) are also displayed, and have functional forms which arise directly from \eqref{eq:Trans_modes} with the leading order expansions \eqref{eq:g_function_2}. Explicitly, one finds
\begin{equation}
 \omega_{0, \perp}^{\mathrm{disk}} \simeq \omega_\mathrm{p} \sqrt{\frac{L}{4 \pi a} \left[ 6 \ln{2} - 1 - 
 2 \ln{\left( \frac{L}{a} \right) }\right]}, \qquad L/a \ll 1, 
 \label{eq:Trans_limits_disk} 
 \end{equation}
  \begin{equation}
 \omega_{0, \perp}^{\mathrm{wire}} \simeq \frac{\omega_\mathrm{p}}{\sqrt{2}} \left( 1 - \frac{4 a}{3 \pi L} \right), \qquad L/a \gg 1. \label{eq:Trans_limits_wire}
 \end{equation}

\subsection{Discussion: comparison to spheroids, screening effects, and damping rates of the plasmonic resonances}
In appendix \ref{Sec:Comparison} we compare our analytical results \eqref{eq:dimer_spectrum_1} and \eqref{eq:Trans_modes} 
for the LSP resonance frequencies of a cylindrical NP to the closed-form expressions for a spheroidal particle with 
the same aspect ratio (see, e.g., the textbook of reference \cite{Bohren1983}) and find an excellent agreement. Such a correspondence 
between both geometries as been previously pointed out by Venermo and Sihvola \cite{Venermo2004}, who compared the polarizability of
a cylinder calculated by means of numerical simulations to that of a spheroid, which is known analytically \cite{Bohren1983}. 
The comparison presented in appendix \ref{Sec:Comparison} thus confirms the relevance as well as the adequacy of our approach, which provides an analytical understanding of plasmonic modes for the cylinder geometry. 

Thus far, our approach has neglected the possible dielectric screening of the valence electrons by the $d$ electrons (characterized by a dielectric constant $\epsilon_\mathrm{d}$), which is of relevance for noble metal NPs, as well as the presence of a dielectric embedding medium (with constant $\epsilon_\mathrm{m}$). 
For the sphere geometry, the presence of screening and the resulting dielectric mismatch notoriously renormalizes 
\cite{Kreibig1995}
the Mie frequency from $\omega_\mathrm{p}/\sqrt{3}$ to $\omega_\mathrm{p}/(\epsilon_\mathrm{d}+2\epsilon_\mathrm{m})^{1/2}$. Within our theoretical approach, it is 
straightforward to realize that when $\epsilon_\mathrm{d}\approx\epsilon_\mathrm{m}=\epsilon$, since the Hartree energy \eqref{eq:Hartree_alpha} is renormalized by a factor $\epsilon^{-1}$, the resonance frequencies in \eqref{eq:dimer_spectrum_1} and \eqref{eq:Trans_modes} take on the same expressions, up to a replacement of $\omega_\mathrm{p}$ by $\omega_\mathrm{p}/\sqrt{\epsilon}$, leading to a redshift of the resonances. The case 
$\epsilon_\mathrm{d}\neq\epsilon_\mathrm{m}$ is much more involved due to the complicated form of the Coulomb interaction in cylindrical coordinates, even within the wire limit \cite{slach06_PRB}, and is out of the scope of the present work. 

A final comment is here in order about the damping of the plasmonic excitations which we have elucidated thus far. Metallic 
nano-objects are subject to radiative and nonradiative damping mechanisms which broaden the resonance of the collective excitation, such that the total decay rate of the LSP modes are given by $\gamma_\alpha=\gamma_\alpha^\mathrm{r}+\gamma^\mathrm{nr}$ ($\alpha=\parallel,\perp$). 
Within our dipolar approximation, the radiative decay rates $\gamma_\alpha^\mathrm{r}$ can be readily estimated from 
the electromagnetic field generated in the far field by a point dipole \cite{Jackson1998} carrying a charge $-eN_\mathrm{e}$ and oscillating at the 
LSP resonance frequency $\omega_{0, \alpha}$. Evaluating the total power radiated by the dipole and the energy initially stored in it, we find
\begin{equation}
\label{eq:gamma_rad}
\gamma_\alpha^\mathrm{r}=\frac{\omega_\mathrm{p}^2\omega_{0, \alpha}^2}{6c^3}a^2L, 
\end{equation} 
with $c$ the speed of light in vacuum. The radiative damping rates $\gamma_\alpha^\mathrm{r}$ thus depend on the cylinder 
dimensions through the explicit dependence $a^2L$ displayed by the equation above, but also through the aspect-ratio dependence of $\omega_{0, \alpha}$ (see figure \ref{fig:frequencies_cylinder}), and increases with the dimensions of the cylinder. 
Using the expansions \eqref{eq:disk1}, \eqref{eq:wire1}, \eqref{eq:Trans_limits_disk}, and \eqref{eq:Trans_limits_wire}, 
we find for the longitudinal mode $\gamma_\parallel^\mathrm{r,disk}\simeq\omega_\mathrm{p}^4a^2L/6c^3$ in the disk limit 
and $\gamma_\parallel^\mathrm{r,wire}\simeq4\omega_\mathrm{p}^4a^3/9\pi c^3$ in the wire limit, which, 
interestingly, does not depend on $L$, since $\omega_{0,\parallel}^\mathrm{wire}$ goes to zero for $L/a\gg1$ [see \eqref{eq:wire1}].
For the transverse mode we find $\gamma_\perp^\mathrm{r,disk}\simeq(6\ln{2}-1)\omega_\mathrm{p}^4aL^2/24\pi c^3$
and $\gamma_\perp^\mathrm{r,wire}\simeq\omega_\mathrm{p}^4a^2L/12c^3$. 

The nonradiative contribution $\gamma^\mathrm{nr}=\gamma_\mathrm{O}+\gamma_\mathrm{L}$ to the total LSP linewidth, which is mode-independent in a first approximation, can be divided into two parts. The first part corresponds to the Ohmic, bulk-like contribution $\gamma_\mathrm{O}$ which essentially arises from electron-phonon and electron-electron scattering. 
The experiments on single gold nanorods protected by a silica shell of reference \cite{Juve2013} report a 
value $\gamma_\mathrm{O}\approx\unit[65]{meV/\hbar}$.
The second part is the Landau 
damping decay rate $\gamma_\mathrm{L}$, a purely quantum-mechanical effect \cite{Brack1993, Bertsch1994} which comes from the confinement of the electronic eigenstates within the NP, and which reads $\gamma_\mathrm{L}=Av_\mathrm{F}/\ell_\mathrm{eff}$, with $A$ a (material and dielectric environment-dependent) constant 
of order $1$, $v_\mathrm{F}$ the Fermi velocity, and $\ell_\mathrm{eff}$ an effective confinement length. The experiments of reference \cite{Juve2013} have shown that $\ell_\mathrm{eff}=(aL)^{1/2}$ provides a good fit to the 
measured data. In these experiments on individual gold nanorods having lengths $L$ in between $\unit[32]{nm}$ and $\unit[70]{nm}$ and radii $a$ in the range $\unit[4.3]{nm}$ to $\unit[11]{nm}$, Landau damping was 
shown to largely dominate the size-dependent part of the total linewidth (which is in the $80$--$\unit[140]{meV/\hbar}$ range), while the 
maximal value of the radiative damping decay rate reported is only $\unit[15]{meV/\hbar}$.

\section{Frequency renormalization due to the spill-out effect}
\label{Sec:single_cylinder_spill}

So far, our approach has been purely classical, and has neglected the spill out of the electronic wave functions 
outside of the NP. This approximation follows from our assumed hard wall mean-field potential, resulting in the approximate density of valence electrons given by \eqref{eq:Long_Hartree}. However, the quantum-mechanical spill-out effect is known to renormalize the LSP resonance frequencies, and is particularly prominent for NPs of only a few nanometers in size~\cite{Brack1993}. 
We thus relax the above hard-wall approximation, and assume that the mean-field potential (including both the ionic positive background and the electron-electron interactions) seen by the valence electrons of the NP is given by 
\begin{equation}
\label{eq:mean-field}
V(\mathbf{r}) = V_0\left[1-\Theta \left(a-r\right) \Theta \left(\frac{L}{2}-|z|\right)\right], 
\end{equation}
where $V_0=\epsilon_\mathrm{F}+W$ is the height of the potential, with $\epsilon_\mathrm{F}$ and $W$ the Fermi energy and the work function of the NP, respectively. 
Such a hypothesis has been tested using density functional ab initio calculations using the local density approximation in reference \cite{Weick2005}, and is a fairly good approximation to the realistic mean-field potential. 

Due to the finite height $V_0$ of the mean-field potential \eqref{eq:mean-field}, some part of the valence electrons can spill out of the cylindrical NP, effectively increasing its length and radius according to the replacements
\begin{equation}
\label{eq:other_dimensions}
 L\rightarrow \tilde{L} = L + 2\ell_\parallel, \qquad\qquad
  a\rightarrow\tilde{a} = a + \ell_\perp.
 \end{equation}
Here, the small spill-out lengths $\ell_\parallel\ll L$ and $\ell_\perp\ll a$ in the longitudinal ($\hat z$) and transverse ($\hat r$) directions, respectively, can be estimated from the average number of spill-out electrons $\mathcal{N}_\parallel$ and $\mathcal{N}_\perp$
in both of these directions according to
\begin{equation}
\label{eq:Cylinder_Spill_both_4}
\ell_\parallel =   \frac 12\frac{\mathcal{N}_\parallel}{N_\mathrm{e}} L, \qquad\qquad
\ell_\perp = \frac{1}{2}   \frac{\mathcal{N}_\perp}{N_\mathrm{e}} a.
 \end{equation}
In the following, we will estimate $\mathcal{N}_\parallel$ and $\mathcal{N}_\perp$ using semiclassical expansions, which will give us access to the spill-out lengths $\ell_\parallel$ and $\ell_\perp$. We will then incorporate the prescription \eqref{eq:other_dimensions} into the mode frequencies \eqref{eq:dimer_spectrum_1} and \eqref{eq:Trans_modes}, which will then provide us with an estimate of the renormalized resonance frequencies.

\subsection{Average number of spill-out electrons and spill-out lengths}

At zero temperature, the average number of spill-out electrons in the longitudinal and transverse directions are given by
\begin{equation}
\label{eq:N_out_def}
\mathcal{N}_\parallel=\sum_\lambda^{\mathrm{occ}}
\int_{\substack{r<a\\|z|>L/2}}\mathrm{d}^3\mathbf{r}\,\left|\psi_\lambda(\mathbf{r})\right|^2,\qquad\qquad
\mathcal{N}_\perp=\sum_\lambda^{\mathrm{occ}}
\int_{\substack{r>a\\|z|<L/2}}\mathrm{d}^3\mathbf{r}\,\left|\psi_\lambda(\mathbf{r})\right|^2,
\end{equation}
respectively. Here, $\lambda$ labels the bound states in the mean-field potential \eqref{eq:mean-field} and the summations run over occupied states up to the Fermi level. The single-particle wave function $\psi_\lambda(\mathbf{r})$ obeys the time-independent Schr\"odinger equation
\begin{equation}
\label{eq:Sch}
\left[-\frac{\hbar^2}{2m_\mathrm{e}}\nabla^2+V(\mathbf{r})\right]\psi_\lambda(\mathbf{r})=\epsilon_\lambda\psi_\lambda(\mathbf{r}), 
\end{equation}
with $\epsilon_\lambda$ the corresponding eigenenergies. Note that in \eqref{eq:N_out_def}, we disregard the negligible number of spill-out electrons arising at the corners of the cylindrical NP. 

The choice of mean-field potential \eqref{eq:mean-field} leads to a nonseparable Schr\"odinger equation \eqref{eq:Sch}. However, the replacement
\begin{equation}
\label{eq:mean-field_approx}
V(\mathbf{r}) \simeq V_0\left[\Theta \left(r-a\right) +\Theta \left(|z| - \frac{L}{2}\right)\right]
\end{equation}
is both an excellent approximation for the original $V(\mathbf{r})$, with only the corners of the cylinder deviating from the nonseparable potential \eqref{eq:mean-field}, and leads to an exactly solvable problem.
Decomposing the separable potential \eqref{eq:mean-field_approx} into $V(\mathbf{r})=V_r(r)+V_z(z)$ with $V_r(r) = V_0 \Theta (r-a)$ and $V_z(z) = V_0 \Theta (|z| - L/2)$, the stationary Schr\"odinger equation \eqref{eq:Sch} then reads 
\begin{equation}
\label{eq:Full_SE}
 \left\{\frac{\partial^2}{\partial r^2}  + \frac{1}{r}\frac{\partial}{\partial r}   + \frac{1}{r^2} \frac{\partial^2}{\partial \theta^2} 
 + \frac{\partial^2}{\partial z^2} + k^2 - \frac{2 m_\mathrm{e}}{\hbar^2} \big[ V_r(r) + V_z(z) \big]\right\} \psi_{n m \tilde{n}}(\mathbf{r}) = 0,
\end{equation}
where $k = \sqrt{2 m_\mathrm{e} \epsilon / \hbar^2}$, and 
where $m$ is the magnetic quantum number and $n$ ($\tilde{n}$) is the principal quantum number due to the transverse (longitudinal) motion. 
We separate the variables in \eqref{eq:Full_SE} using 
\begin{equation}
\label{eq:psi_sep}
\psi_{n m \tilde{n}} (\mathbf{r}) = F_{n m}(r, \theta) Z_{\tilde{n}}(z),
\end{equation}
 and are thus led to two Schr{\"o}dinger equations in the reduced eigenvalues $k_{r }$ and $k_z$, respectively, where $k^2 =  k_{r}^2+k_z^2$, whose solutions are given explicitly in appendix \ref{sec:app}. 
 Using the results presented there [see in particular \eqref{eq:electron_integrals_3}], we are then able to evaluate 
 the integrals entering \eqref{eq:N_out_def}, which are approximately given in the high-energy, semiclassical limit of $k_0a\gg1$ and $k_0L\gg1$ [with $k_0=(2m_\mathrm{e}V_0/\hbar^2)^{1/2}$] by
\begin{equation}
 \label{eq:electronic_int_approx}
\int_{\substack{r<a\\|z|>L/2}}\mathrm{d}^3\mathbf{r}\,\left|\psi_{nm\tilde n}(\mathbf{r})\right|^2
\simeq  \frac{2}{\kappa_{z} L} \frac{k_z^2}{k_0^2} , \qquad\qquad
\int_{\substack{r>a\\|z|<L/2}}\mathrm{d}^3\mathbf{r}\,\left|\psi_{nm\tilde n}(\mathbf{r})\right|^2\simeq
\frac{1}{2\kappa_{r} a},
 \end{equation}
where $\kappa_{r}=(k_0^2-k_{r})^{1/2}$ and $\kappa_{z}=(k_0^2-k_{z})^{1/2}$.

Upon substituting the expressions \eqref{eq:electronic_int_approx} into \eqref{eq:N_out_def}, we then replace the summation over the set of quantum numbers $n$, $m$, and $\tilde n$ by an integral over wave vector 
$\mathbf{k}$. We take for the density of states the 
leading-in-$\hbar$ Weyl term \cite{brack}, which is appropriate in the 
semiclassical limit $k_\mathrm{F}a\gg1$ and $k_\mathrm{F}L\gg1$ (with $k_\mathrm{F}$ the Fermi wave vector). For typical noble metals, 
such as, e.g., Ag or Au, one has $k_\mathrm{F}a\simeq 10\,a[\mathrm{nm}]$, so that the semiclassical approximation is suitable even for 
nanometer-sized NPs \cite{Weick2006}.
 The aforementioned prescription leads to
\begin{equation}
\mathcal{N}_\parallel\simeq 2\frac{V}{(2\pi)^3}\int_{k<k_\mathrm{F}}\mathrm{d}^3\mathbf{k}\,\frac{2}{\kappa_{z} L} \frac{k_z^2}{k_0^2}
,\qquad\qquad
\mathcal{N}_\perp\simeq 2\frac{V}{(2\pi)^3}\int_{k<k_\mathrm{F}}\mathrm{d}^3\mathbf{k}\,\frac{1}{2\kappa_{r } a},
\end{equation}
where the prefactor of $2$ accounts for the spin degeneracy and $V=\pi a^2L$ is the volume of the cylinder. 
Performing the above integrals in spherical coordinates, we arrive at 
\begin{equation}
\label{eq:Cylinder_Spill_both}
\mathcal{N}_\parallel = \frac{k_0^2 a^2}{\pi} \int_0^{{k_\mathrm{F}}/{k_0}} \mathrm{d}x\; x^4 
 \int_{-1}^{+1} \mathrm{d}t\; \frac{t^2}{\sqrt{1-x^2 t^2}} , \quad
\mathcal{N}_\perp = \frac{k_0^2 a L}{4 \pi} \int_0^{{k_\mathrm{F}}/{k_0}} \mathrm{d}x\; x^2  
\int_{-1}^{+1} \mathrm{d}t\; \frac{1}{\sqrt{1-x^2 (1-t^2)}}, 
 \end{equation}
where $k_0 > k_\mathrm{F}$, and where the integrals with respect to the dimensionless radial ($x$) and polar ($t$) coordinates are yet to be performed. 
Evaluating the above integrals \eqref{eq:Cylinder_Spill_both}, we find the expressions
\begin{equation}
\label{eq:Cylinder_Spill_both_2}
\mathcal{N}_\parallel = \frac{(k_\mathrm{F} a)^2}{4 \pi} h_\parallel\left( \frac{\epsilon_\mathrm{F}}{V_0} \right), \qquad\qquad
\mathcal{N}_\perp = \frac{k_\mathrm{F}^2 a L}{4 \pi} h_\perp\left( \frac{\epsilon_\mathrm{F}}{V_0} \right). 
\end{equation}
Here we have introduced the auxiliary functions
\begin{equation}
\label{eq:other_fun}
  h_\parallel(x) = \left(\frac 32 - x\right) \sqrt{\frac{1}{x}-1} + \left(2-\frac{3}{2x}\right) \arcsin{\sqrt{x}}, \qquad
  h_\perp(x) = \frac{1}{\sqrt{x}} + \left(1-\frac 1x\right) \arctanh{\sqrt{x}}.
 \end{equation}
Scaling the results \eqref{eq:Cylinder_Spill_both_2} with the total number of electrons in the NP $N_\mathrm{e} = L a^2 k_\mathrm{F}^3/3 \pi$, we obtain
\begin{equation}
\label{eq:Cylinder_Spill_both_3}
  \frac{\mathcal{N}_\parallel}{N_\mathrm{e}} = \frac{3}{4 k_\mathrm{F} L} h_\parallel\left( \frac{\epsilon_\mathrm{F}}{V_0} \right),  \qquad\qquad
  \frac{\mathcal{N}_\perp}{N_\mathrm{e}} = \frac{3}{4 k_\mathrm{F} a} h_\perp\left( \frac{\epsilon_\mathrm{F}}{V_0} \right).
 \end{equation}
Thus, the fraction of spill-out electrons in both the longitudinal and transverse directions scale with the inverse of the
spatial extent of the cylinder ($\propto 1/a$, $1/L$), and so becomes increasingly important for particles with nanometric dimensions. 

\begin{figure*}[tb]
 \includegraphics[width=1.0\linewidth]{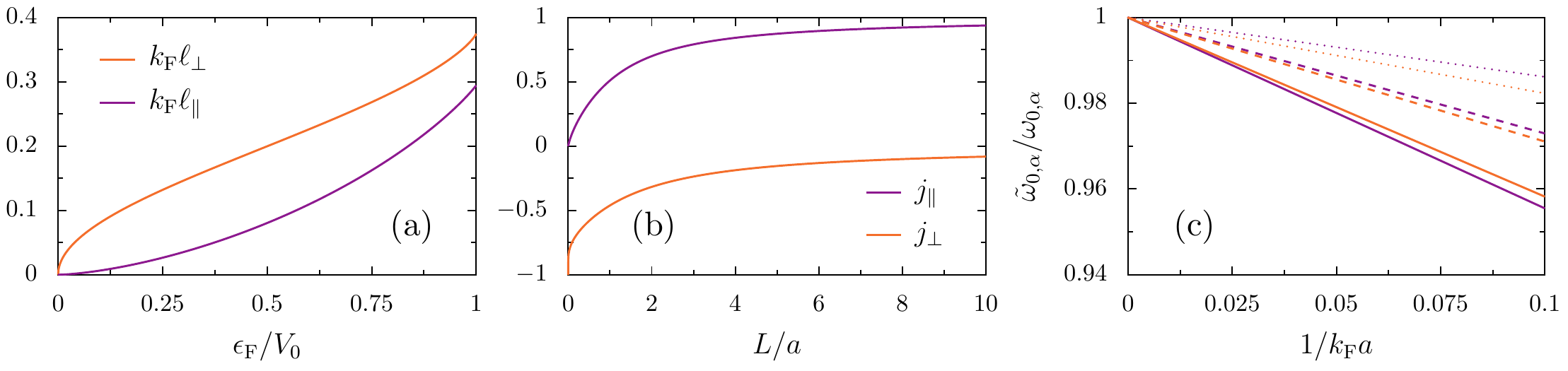}
 \caption{(a) Spill-out lengths \eqref{eq:Cylinder_Spill_both_5}, scaled with the Fermi wave vector $k_\mathrm{F}$, as a function of the ratio of the Fermi energy to mean-field potential strength, $\epsilon_\mathrm{F}/V_0$.
 (b) Auxiliary functions $j_\alpha$ ($\alpha=\parallel,\perp$) from \eqref{eq:j} as a function of the aspect ratio $L/a$.
 (c) Renormalized resonance frequencies $\tilde\omega_{0,\parallel}$ (violet lines) and $\tilde\omega_{0,\perp}$ (orange lines) 
 from \eqref{eq:omega_tilde} in units of the bare frequencies $\omega_{0,\alpha}$ as a function of the inverse size of the nanoparticle 
 (dotted lines: $\epsilon_\mathrm{F}/V_0=0.25$; dashed lines: $\epsilon_\mathrm{F}/V_0=0.50$; solid lines: $\epsilon_\mathrm{F}/V_0=0.75$), for $L/a=1$.} 
 \label{fig:spillout}
\end{figure*}

With the above results \eqref{eq:Cylinder_Spill_both_3}, we can now evaluate the spill-out lengths \eqref{eq:Cylinder_Spill_both_4}, which read
\begin{equation}
\label{eq:Cylinder_Spill_both_5}
    k_\mathrm{F} \ell_\parallel = \frac{3}{8} h_\parallel\left( \frac{\epsilon_\mathrm{F}}{V_0} \right), \qquad\qquad
    k_\mathrm{F} \ell_\perp = \frac{3}{8} h_\perp\left( \frac{\epsilon_\mathrm{F}}{V_0} \right). 
 \end{equation}
Importantly, these two quantities do not depend on the NP dimensions $L$ and $a$, and only on the 
Fermi energy $\epsilon_\mathrm{F}$ (or the Fermi wave vector $k_\mathrm{F}$) and the depth $V_0$ of the mean-field potential \eqref{eq:mean-field}. 
The spill-out lengths \eqref{eq:Cylinder_Spill_both_5} are plotted
in figure \ref{fig:spillout}(a) as a function of $\epsilon_\mathrm{F}/V_0$. As one can see from the figure, both of these quantities smoothly increase with the above-mentioned ratio. Since $k_\mathrm{F}$ is typically of the order of 
$\unit[10^{8}]{cm^{-1}}$ for alkaline or noble metals, and since $\epsilon_\mathrm{F}/V_0$ is roughly of the order of $0.5$ \cite{Brack1993}, the spill-out lengths \eqref{eq:Cylinder_Spill_both_5} are only of a few tenths of an angstrom. However, as we will see in the next section, such a tiny spread of the electronic wave functions outside of the NP may have a non-negligible effect on the LSP resonance frequency.

\subsection{Frequency redshifts due to the spill-out effect}
\label{Sec:single_cylinder_spillout}

We are now in a position to calculate the renormalized resonance frequency in the longitudinal (transverse) polarization $\tilde\omega_{0,\parallel}$ ($\tilde\omega_{0,\perp}$) due to the spill-out effect.
We account for the spill-out of the electrons by treating the cylindrical NP with the effective dimensions 
$\tilde{L}$ and $\tilde{a}$ as in \eqref{eq:other_dimensions}. 
It follows from the direct substitution of these effective dimensions into the resonance frequencies \eqref{eq:dimer_spectrum_1} and \eqref{eq:Trans_modes}, which assumed hard-wall confinement of the valence electrons, that the renormalized resonance frequencies are, to leading order in the scaled spill-out lengths $\ell_\parallel/L$ and $\ell_\perp/a$ [cf.\ \eqref{eq:Cylinder_Spill_both_5}], given by 
\begin{equation}
\label{eq:omega_tilde}
\tilde\omega_{0,\alpha}\simeq\omega_{0,\alpha}\left\{
1-\left[1+j_\alpha\left(\frac La\right)\right]\frac{\ell_\parallel}{L}
-\left[1-\frac 12 j_\alpha\left(\frac La\right)\right]\frac{\ell_\perp}{a}\right\},\qquad\alpha=\parallel, \perp.
\end{equation}
Here, we defined the two functions
\begin{align}
\label{eq:j}
j_\parallel(x)=\frac{(2/\pi x)\left[4/3-g(x)+xg'(x)\right]}{1+(2/\pi x)[4/3-g(x)]},\qquad\qquad
j_\perp(x)=\frac{4/3-g(x)+xg'(x)}{4/3-g(x)},
\end{align}
where 
\begin{equation}
g'(x)=\frac 12\left[\left(x^2+4\right){K}\left(-\frac{4}{x^2}\right)
-x^2{E}\left(-\frac{4}{x^2}\right)\right]
\end{equation}
is the derivative with respect to $x$ of the function $g(x)$ defined in \eqref{eq:g_function}.

We plot in figure \ref{fig:spillout}(b)  the auxiliary functions \eqref{eq:j}, which are both monotonically increasing functions of the aspect ratio $L/a$ of the cylinder sketched in the inset of figure \ref{fig:frequencies_cylinder}, with $0< j_\parallel<1$ and $-1< j_\perp<0$. Thus, the prefactors of the terms $\propto 1/L$ and $\propto 1/a$ in \eqref{eq:omega_tilde} are negative, such that the LSP resonance frequencies of the cylinder experience a redshift  as the NP size decreases, as it is the case for the sphere geometry \cite{Brack1993}. This is exemplified in figure \ref{fig:spillout}(c), which displays 
the renormalized resonance frequencies \eqref{eq:omega_tilde} (in units of the bare ones) as a function of $1/k_\mathrm{F}a$, for an aspect ratio
$L/a=1$, and for increasing values of $\epsilon_\mathrm{F}/V_0$. As one can see from the figure, the deviation due to the spill-out effect from the bare resonance frequencies can reach between ca.\ $\unit[1.5]{\%}$ to $\unit[4.5]{\%}$ (depending on the ratio $\epsilon_\mathrm{F}/V_0$, which is essentially material-dependent) for $k_\mathrm{F}a=10$ (which typically corresponds to $a\simeq\unit[1]{nm}$ for normal metals). 
Progress in nanofabrication techniques should allow one to vary the cylindrical NP size and thus to measure the predicted size dependence of the resonance frequencies~\eqref{eq:omega_tilde}.

\section{Coupled modes in a cylindrical dimer}
\label{Sec:dimer}

\begin{figure}[tb]
\begin{center}
 \includegraphics[width=0.5\linewidth]{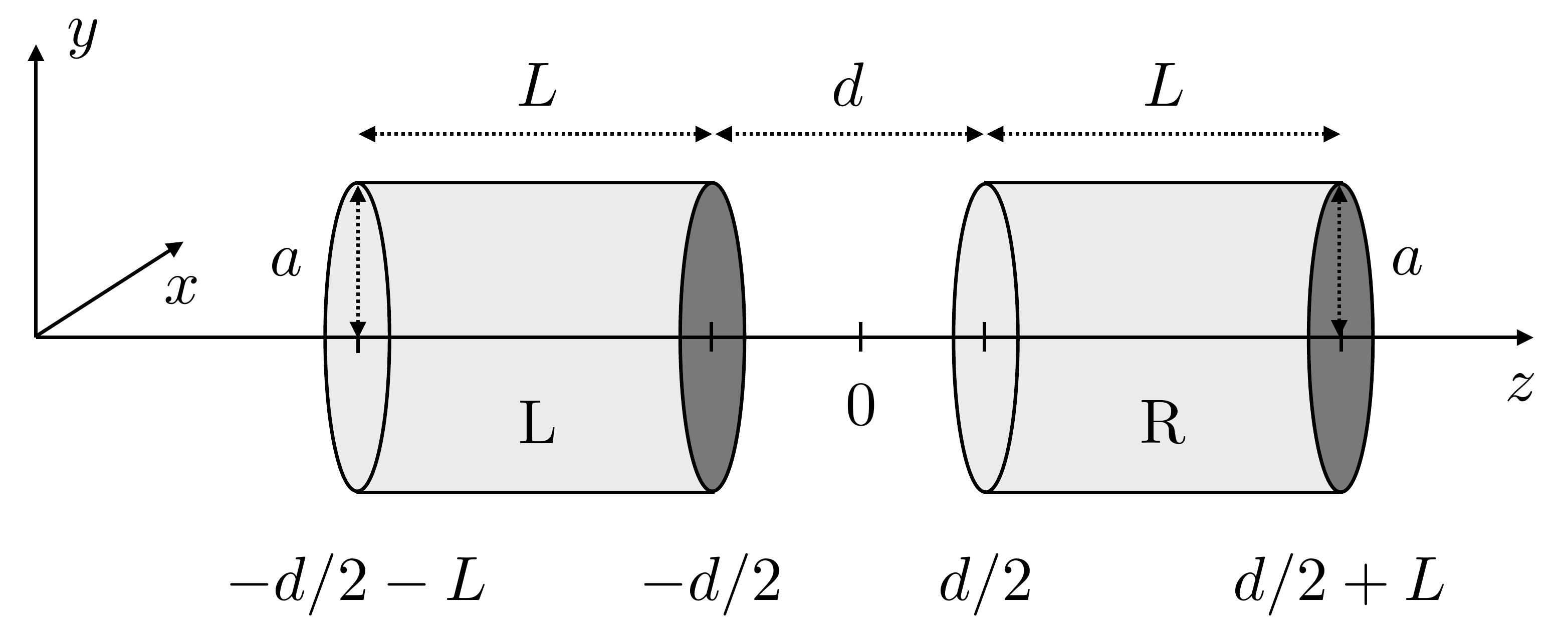}
 \caption{Sketch of a dimer of cylindrical metallic nanoparticles, denoted $\mathrm{L}$ and $\mathrm{R}$. Each nanoparticle is of radius $a$ and length $L$, and they are separated by an end-to-end gap distance $d$.}
 \label{fig:sketch}
 \end{center}
\end{figure}

We now consider a dimer of cylindrical NPs composed of the same metal, which are aligned along the ${z}$ axis, and are separated by a distance $d$ (see figure \ref{fig:sketch}). The two cylinders, denoted left ($\mathrm{L}$) and right ($\mathrm{R}$), are both of a radius $a$ and length $L$, and contain $N_\mathrm{e}$ valence electrons each. In the following, we study the resultant coupled plasmonic modes in the dimer, for both the longitudinal ($\alpha=\parallel$) and transverse ($\alpha=\perp$) polarizations, in a similar fashion to the single cylinder calculation of \S\ref{sec:1_cyl}.

Notably, we focus on the dipolar modes and do not take into account higher-order contributions to the collective plasmon-plasmon interactions, which are only relevant for small interparticle separations.\footnote{For spheres, this non-dipolar regime is given by $s \lesssim 3 R$, where $R$ is the NP radius and $s$ is the point-to-point separation~\cite{Park2004}.} We also focus on separation distances such that 
the near-field coupling between the LSPs on each NP dominates, and thus
disregard retardation effects. The latter only lead to small frequency renormalization effects in NP dimers~\cite{DowningLamb2017} (although retardation effects can be significant in long NP chains~\cite{Weber2004, DowningCollectiveLamb2017, DowningLamb2018}).

We start by neglecting the spill-out of the electronic wave functions outside of each cylinder. As in \S\ref{sec:1_cyl}, we assume that the density $n(\mathbf{r})$ of valence electrons is uniform within each cylinder (with density $n_0$), and vanishing outside. We thus have
\begin{equation}
\label{eq:density}
 n(\mathbf{r}) = n_\mathrm{L} (\mathbf{r}) + n_\mathrm{R} (\mathbf{r}), 
\end{equation}
where, in cylindrical coordinates, the single NP contributions are
\begin{subequations}
\label{eq:density_left_right}
\begin{align}
 n_\mathrm{L} (\mathbf{r}) &= n_0\, \Theta \left(a-r\right)  \Theta \left(-z-\frac{d}{2}\right) \Theta \left(L+\frac{d}{2}+z\right), \\
 n_\mathrm{R} (\mathbf{r}) &= n_0\, \Theta \left(a-r\right)  \Theta \left(z-\frac{d}{2}\right) \Theta \left(L+\frac{d}{2}-z\right).
 \end{align}
\end{subequations}  
In order to have access to the frequencies of the coupled dipolar modes, we shall impose two types of rigid 
displacements $\mathbf{u}_{\alpha}^\tau$ on the total density \eqref{eq:density} for each polarization $\alpha$, distinguished by the index $\tau=\pm$. To first order in the displacement field $\mathbf{u}^\tau_{\alpha}$, the electronic density is perturbed like $n(\mathbf{r}-\mathbf{u}_{\alpha}^\tau) \simeq n(\mathbf{r}) + \delta n_{\alpha}^\tau(\mathbf{r})$, where
\begin{equation}
\label{eq:density_delta}
\delta n_{\alpha}^\tau  (\mathbf{r}) = - \mathbf{u}_{\alpha}^\tau \cdot \nabla n (\mathbf{r}).
\end{equation}
This density change gives rise to the two possible normal modes of the system (labelled by $\tau$) for each polarization $\alpha$, which we now consider in turn.

\subsection{Longitudinal modes}
\label{Sec:dimer_long}

Firstly, let us consider the symmetric mode in the longitudinal polarization ($\rightarrow \rightarrow$), which we denote with the index $\tau = -$ since it constitutes the low-energy mode. We enforce the displacement $\mathbf{u}_{\parallel}^-  = u\, \hat{z}$ in both cylinders, such that the shift in density \eqref{eq:density_delta} is
$\delta n_{\parallel}^-(\mathbf{r}) = \delta n_{\mathrm{L}, \parallel}(\mathbf{r}) + \delta n_{\mathrm{R}, \parallel}(\mathbf{r})$, 
with
\begin{subequations}
\label{eq:_change_density_left_right}
\begin{align}
 \delta n_{\mathrm{L}, \parallel} (\mathbf{r}) &= u n_0\, \Theta \left(a-r\right) \left[ \delta \left(z+\frac{d}{2}\right) - \delta \left(L+\frac{d}{2}+z\right) \right], \\
 \delta n_{\mathrm{R}, \parallel} (\mathbf{r}) &= u n_0\, \Theta \left(a-r\right) \left[ \delta \left(L+\frac{d}{2}-z\right) - \delta \left(z-\frac{d}{2}\right) \right].
 \end{align}
\end{subequations} 
Secondly, let us consider the antisymmetric, high-energy mode ($\rightarrow \leftarrow$), which we mark with the index $\tau = +$. Imposing the rigid displacements $\mathbf{u}_{\parallel}^+  = \pm u\, \hat{z}$, which act in different directions ($\pm$) in each cylinder ($\mathrm{L}$ relative to $\mathrm{R}$), the change in density \eqref{eq:density_delta} reads
$\delta n_{\parallel}^+(\mathbf{r}) = \delta n_{\mathrm{R}, \parallel}(\mathbf{r}) - \delta n_{\mathrm{L}, \parallel}(\mathbf{r})$, 
in terms of the quantities~\eqref{eq:_change_density_left_right}.

The change in electrostatic interactions between the electronic clouds is described by the change in the Hartree energy
\begin{equation}
\label{eq:change_Hartee}
 \delta E^{\tau}_{\parallel} = \delta E^{\mathrm{LL}}_{\parallel} + \delta E^{\mathrm{RR}}_{\parallel} - \tau \left( \delta E^{\mathrm{LR}}_{\parallel} 
 + \delta E^{\mathrm{RL}}_{\parallel} \right),
\end{equation}
where we have used the decomposition
\begin{equation}
\label{eq:change_Hartee_2}
 \delta E^{ij}_{\alpha} = \frac{e^2}{2} \int \mathrm{d}^3\mathbf{r} \int\mathrm{d}^3\mathbf{r'} \frac{\delta n_{i, \alpha} (\mathbf{r}) \delta n_{j, \alpha} (\mathbf{r'})}{|\mathbf{r}-\mathbf{r'}|},
\end{equation}
where $i, j \in \{ \mathrm{L}, \mathrm{R} \}$, $\alpha \in \{ \parallel, \perp \}$ and $\delta n_{i, \parallel} (\mathbf{r})$ are given in 
\eqref{eq:_change_density_left_right}. The first two terms on the right-hand side of \eqref{eq:change_Hartee} describe the contributions from each cylinder in isolation, 
$\delta E^{\mathrm{LL}}_{\parallel} = \delta E^{\mathrm{RR}}_{ \parallel} = \delta E_\parallel$,
where $\delta E_\parallel$ is given in \eqref{eq:Long_Elliptic}. The effect of electrostatic coupling is contained in the final two terms of \eqref{eq:change_Hartee}. An analogous calculation to that which led to \eqref{eq:Long_Elliptic} yields for the remaining two terms in \eqref{eq:change_Hartee} the expression
\begin{equation}
\label{eq:Hartee_2}
 \delta E^{\mathrm{LR}}_{\parallel} = \delta E^{\mathrm{RL}}_{\parallel} = 2 \pi \left( e u n_0 \right)^2 a^3 
 \left[ 2g\left(\frac{L+d}{a}\right) -  g\left(\frac{d}{a}\right) -  g\left(\frac{2 L+d}{a}\right) \right], 
\end{equation}
where the function $g(x)$ is defined in \eqref{eq:g_function}.

The restoring force $F_{\alpha}^\tau = - \partial_u (\delta E_\alpha^\tau)=- k_{\alpha}^\tau u$ then provides us with a relation to the effective spring constant $k_{\alpha}^\tau$ of the problem, from which the resonance frequencies $\omega_{\tau, \alpha} = \sqrt{k_{\alpha}^\tau / 2 N_\mathrm{e} m_\mathrm{e}}$ are revealed.\footnote{Note that $2 N_\mathrm{e} m_\mathrm{e}$ corresponds to the total electronic mass of the dimer.} 
Hence the longitudinal eigenfrequencies of a dimer of cylindrical NPs are given by
\begin{equation}
\label{eq:dimer_spectrum}
\omega_{\tau, \parallel} = \sqrt{\omega_{0,\parallel}^2 + 2 \tau \Omega^2},   
\end{equation}
which accounts for both the high-energy ($\tau = +, \rightarrow \leftarrow$) and the low-energy ($\tau = -, \rightarrow \rightarrow$) modes. Here, $\omega_{0,\parallel}$
is the resonance frequency of an isolated cylinder \eqref{eq:dimer_spectrum_1}, which depends on the aspect ratio $L/a$, while 
the effective coupling constant of the problem $\Omega$ has the functional form
\begin{equation}
\label{eq:dimer_coupling}
\Omega = \omega_\mathrm{p} \sqrt{ \frac{a}{2 \pi L} \left[ g \left( \frac{d}{a} \right) +  g \left( \frac{2 L + d}{a} \right) - 2g \left( \frac{L + d}{a} \right) \right]}
\end{equation}
and depends on both $L/a$ and $d/a$.
Equation \eqref{eq:dimer_spectrum} analytically describes the longitudinal resonance frequencies of the pair of coupled dipolar modes of the cylindrical dimer, provided the separation $d$ is not too small (i.e., where one can neglect multipolar effects and/or the quantum tunneling of electrons between the two NPs comprising the dimer). 

Interestingly, we show in appendix \ref{sec:dipole} that the result \eqref{eq:dimer_spectrum},  
in the limit of large interparticle distance (i.e., $d/a\gg1$ and $d/L\gg1$), for which we have [using \eqref{eq:g_function_2b}]
\begin{equation}
\label{eq:Omega_approx}
\Omega\simeq\omega_\mathrm{p}\sqrt{\frac{La^2}{4d^3}}, 
\end{equation}
can be found from a simple model of coupled point dipoles. 
From \eqref{eq:Omega_approx}
one further immediately notices that in the extreme isolation limit ($d \to \infty$), the coupling constant \eqref{eq:dimer_coupling} vanishes and thus the single NP resonance frequency \eqref{eq:dimer_spectrum_1} is recovered by \eqref{eq:dimer_spectrum}. 

\begin{figure}[tb]
\begin{center}
 \includegraphics[width=1.0\linewidth]{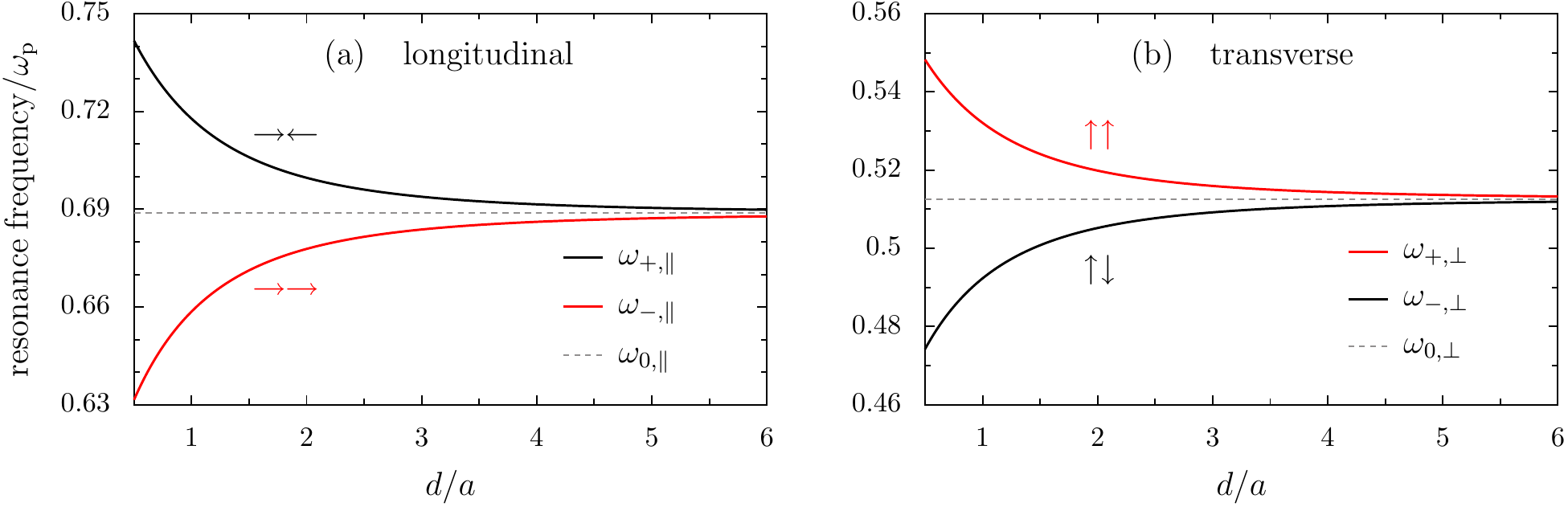}
 \caption{Resonance frequencies of the dark (black lines) and bright (red lines) coupled modes, in units of the plasma frequency $\omega_\mathrm{p}$, as a function of interparticle separation $d$, in units of the cylinder radius $a$. Panel (a): longitudinal modes $\omega_{\tau, \parallel}$, as encapsulated by \eqref{eq:dimer_spectrum}, and the associated single NP result $\omega_{0,\parallel}$ from \ \eqref{eq:dimer_spectrum_1}. Panel (b): transverse modes $\omega_{\tau, \perp}$, as given by \eqref{eq:dimer_spectrum_trans}, and the corresponding single NP result $\omega_{0,\perp}$ from \ \eqref{eq:Trans_modes}. In the figure, the aspect ratio $L/a = 1$.}
 \label{fig:frequencies_dimer}
 \end{center}
\end{figure}

We plot the dimer resonance frequencies \eqref{eq:dimer_spectrum} as a function of the 
interparticle separation $d$ in figure \ref{fig:frequencies_dimer}(a), 
for both the high-energy (black line) and the low-energy (red line) modes, for the aspect ratio $L/a = 1$. 
Evidently, for separations of just a few multiples of the cylinder size, the two resonance frequencies 
converge onto the one of a single cylinder $\omega_{0,\parallel}$ (see the gray dashed line).
 As one can see from the figure, for separations below the cylinder size, the pair of plasmonic levels become increasingly distinct and hence experimentally detectable. The bright mode ($\rightarrow \rightarrow$) may be accessed optically, whereas the dark mode ($\rightarrow \leftarrow$) should be probed via electron energy loss spectroscopy (EELS) \cite{Abajo2010}. 
 
For completeness, we now consider the two limiting cases of equation \eqref{eq:dimer_spectrum}, i.e., the nanodisk ($L/a \ll 1$) and nanowire ($L/a \gg 1$) limits, and for large separation distances ($d/a\gg1$, $d/L\gg1$).
Using \eqref{eq:Omega_approx}, the full eigenspectrum \eqref{eq:dimer_spectrum} of the coupled nanodisks reads
$\omega_{\tau, \parallel}^{\mathrm{disk}} \simeq \omega_{0,\parallel}^{\mathrm{disk}} + \tau \omega_\mathrm{p} \frac{L a^2}{4d^3}$,
where $\omega_{0,\parallel}^{\mathrm{disk}}$ is given in \eqref{eq:disk1}, 
analytically demonstrating the expected inverse cubic decay with separation ($\propto 1/d^3$), which is characteristic of a quasistatic dipolar interaction \cite{Jackson1998}.
In the opposite regime of a dimer of wire-like NPs, and further assuming $L^2 a/d^3 \ll 1$, we have 
$\omega_{\tau, \parallel}^{\mathrm{wire}} \simeq \omega_{0,\parallel}^{\mathrm{wire}} + \tau \frac{\omega_\mathrm{p}}{8} \sqrt{\frac{3 \pi}{2}} ( \frac{\sqrt{L a}}{d})^3$,
where an expression for $ \omega_{0,\parallel}^{\mathrm{wire}}$ can be found in \eqref{eq:wire1}, again showcasing the characteristic scaling typical of the dipole-dipole interaction regime ($\propto 1/d^3$).

\subsection{Transverse modes}
\label{Sec:dimer_trans}

The calculation for the transverse polarized coupled modes  ($\alpha = \perp$) proceeds in a similar manner to that of the previous subsection. The main difference is that the symmetric mode ($\uparrow \uparrow$) is now the high-energy mode and is associated with $\tau = +$, while the antisymmetric mode ($\uparrow \downarrow$) corresponds to the low-energy mode and has the label $\tau = -$.

We again consider two separate displacements in order to characterize the two coupled dipolar modes. Firstly we assume 
$\mathbf{u}_{\perp}^+  = u\, \hat{x}$, which gives rise to the symmetric mode, and secondly we let $\mathbf{u}_{\perp}^-  = \pm u\, \hat{x}$, which distinguishes the antisymmetric mode (in the right hand side of $\mathbf{u}_{\perp}^-$, the $\pm$ refers to the displacements being in opposite directions for cylinder $\mathrm{L}$ as compared to $\mathrm{R}$). The shift in electronic density across the dimer then reads $\delta n_{\perp}^\tau(\mathbf{r}) = \delta n_{\mathrm{L}, \perp}(\mathbf{r}) + \tau \delta n_{\mathrm{R}, \perp}(\mathbf{r})$, where
\begin{subequations}
\label{eq:_change_density_left_right_trans}
\begin{align}
 \delta n_{\mathrm{L}, \perp} (\mathbf{r}) &= u n_0\, \cos{\theta}\, \delta \left(a-r\right) \Theta \left( -z - \frac{d}{2}  \right) \Theta \left( L + \frac{d}{2} + z \right), \\
 \delta n_{\mathrm{R}, \perp} (\mathbf{r}) &= u n_0\, \cos{\theta}\, \delta \left(a-r\right) \Theta \left( z - \frac{d}{2} \right) \Theta \left( L + \frac{d}{2} - z \right),
 \end{align}
\end{subequations} 
as follows from \eqref{eq:density_left_right} and \eqref{eq:density_delta}.

The change in Hartree energy accounts for the transverse electrostatic interactions via 
$\delta E^{\tau}_{\perp} = \delta E^{\mathrm{LL}}_{\perp} + \delta E^{\mathrm{RR}}_{\perp} 
+ \tau ( \delta E^{\mathrm{LR}}_{\perp} + \delta E^{\mathrm{RL}}_{ \perp} )$, where we utilized the decomposition of \eqref{eq:change_Hartee_2}. The first two terms in $\delta E^{\tau}_{\perp}$ are single NP contributions, and so are exactly that of \eqref{eq:Trans_Hartree_2}, explicitly 
$\delta E^{\mathrm{LL}}_{\perp}  = \delta E^{\mathrm{RR}}_{ \perp} = \delta E_{\perp}$. The remaining coupling terms in 
$\delta E^{\tau}_{\perp}$ are also equal by symmetry, and are given by
\begin{equation}
\label{eq:Hartee_2_trans}
 \delta E^{\mathrm{LR}}_{\perp} = \delta E^{\mathrm{RL}}_{\perp} = \pi \left( e u n_0 \right)^2 a^3 
 \left[ g\left(\frac{d}{a}\right) +  g\left(\frac{2 L+d}{a}\right) - 2g\left(\frac{L+d}{a}\right) \right], 
\end{equation}
where $g(x)$ was introduced in \eqref{eq:g_function}. As in the previous subsection, the analytic expression for the transverse resonance frequencies soon follows from the restoring force $F^{\tau}_{\perp}$ (see below \eqref{eq:Hartee_2} for details) as
\begin{equation}
\label{eq:dimer_spectrum_trans}
\omega_{\tau, \perp} = \sqrt{ \omega_{0,\perp}^2 + \tau \Omega^2},   
\end{equation}
where the coupling constant $\Omega$ is defined in \eqref{eq:dimer_coupling}, and with the single cylinder resonance frequency $\omega_{0,\perp}$ of \eqref{eq:Trans_modes}. Notably, compared to the longitudinal result $\omega_{\tau, \parallel}$ of \eqref{eq:dimer_spectrum}, the second term under the square root in \eqref{eq:dimer_spectrum_trans}, the so-called coupling term, is half as large. This property is familiar from the canonical case of a dimer of spherical NPs, where the longitudinal modes are also (approximately) twice as widely split as the transverse modes~\cite{Brandstetter2015, DowningLamb2017}. 
Moreover, we note that in the limit of large interparticle distance, the result \eqref{eq:dimer_spectrum_trans} can be inferred from the coupled point dipole model discussed in appendix \ref{sec:dipole}.

The transverse polarized dimer resonance frequencies \eqref{eq:dimer_spectrum_trans} are plotted as a function of the interparticle separation $d$ in figure \ref{fig:frequencies_dimer}(b), with the aspect ratio fixed at $L/a = 1$. The plot illustrates increasingly large inter-mode splittings for decreasing separations $d$, such that $\omega_{\tau, \perp}$ become significantly detuned from the single cylinder result (dashed gray line). In stark contrast to the longitudinal coupled modes of figure \ref{fig:frequencies_dimer}(a), in panel (b) it is the high-energy mode which strongly couples to light (red line), and as such may be detected straightforwardly by optical means, while the low-energy mode is dark (black line), and as such requires EELS probing techniques.

A remark is now in order about the influence of the spill-out effect onto the coupled mode frequencies shown in figure \ref{fig:frequencies_dimer}. Since the effective coupling constant $\Omega$ of \eqref{eq:dimer_coupling} already represents a rather small correction to the single particle results $\omega_{0,\parallel}$ and $\omega_{0,\perp}$ in \eqref{eq:dimer_spectrum} and \eqref{eq:dimer_spectrum_trans}, respectively, we can safely assume that $\Omega$ is not renormalized by the spill-out lengths \eqref{eq:Cylinder_Spill_both_5}. Therefore, the renormalized coupled mode frequencies are $\tilde\omega_{\tau, \parallel} = (\tilde\omega_{0,\parallel}^2 + 2\tau \Omega^2)^{1/2}$ and $\tilde\omega_{\tau, \perp} = (\tilde\omega_{0,\perp}^2 + \tau \Omega^2)^{1/2}$ respectively, where $\tilde\omega_{0,\parallel}$ and $\tilde\omega_{0,\perp}$ are given in \eqref{eq:omega_tilde}. Henceforth, the only effect of the spill-out electrons is a global frequency shift towards the red end of the spectrum, which does not depend on the interparticle distance $d$. 

We conclude this section by commenting on the linewidth of the coupled LSP modes which we have calculated. 
While the nonradiative part (i.e., Ohmic and Landau damping) does not depend in first approximation on the bright or dark nature of the mode and can be evaluated from the single-particle result discussed in \S\S\ref{sec:1_cyl}(c) \cite{Brandstetter2015}, the radiative part is strongly mode-dependent. Indeed, dark modes in dimers of near-field coupled NPs essentially do not radiate and are immune to radiation damping, while the damping rate of the bright modes is approximately given by twice the result \eqref{eq:gamma_rad} for an individual NP.

\section{Conclusion}
\label{Sec:conc}

We have considered the fundamental problem of characterizing analytically within the quasistatic approximation the longitudinal and transverse dipolar modes supported by a cylindrical NP with an arbitrary aspect ratio. Inspired by the trend of increasing miniaturization, we have derived semiclassical expressions for the spill-out lengths in metallic cylinders, and have shown that this quantum phenomenon can significantly renormalize the plasmonic frequencies. Recent experiments on plasmonic cylinders, including with nanodisks and nanowires, suggests that both the resonance frequency dependence on the cylinder aspect ratio and the quantum size effect we consider may be probed in cutting edge laboratories~\cite{Schmidt2012, Juve2013, Schmidt2014, Krug2014, Wang2017, Movsesyan2018, Schmidt2018}.

We have also developed an analytical theory of a dimer of cylindrical NPs, and have derived a simple expression for the bright and dark mode eigenfrequencies of the collective plasmons, for both the longitudinal and transverse polarizations. These results provide insight into the evolution of the splitting of the coupled modes as a function of the interparticle separation, and act as a benchmark for future experimental and numerical studies of collective plasmonic behavior~\cite{Reinhard2005, Jain2007, Jain2010}.

\enlargethispage{20pt}

\dataccess{This article has no additional data.}

\aucontribute{CAD and GW performed the analytical calculations. CAD wrote a first version of the manuscript, which was finalized by GW. GW supervised the project. Both authors gave final approval for publication and agree to be held accountable for the work performed therein.}

\competing{We declare we have no competing interests.}

\funding{We acknowledge support from Agence Nationale de la Recherche (Grant No.\ ANR-14-CE26-0005 Q-MetaMat).
CAD acknowledges support from the Juan de la Cierva program (MINECO, Spain) and from the Royal Society University Research Fellowship (URF\slash R1\slash 201158).
This research project has been supported by the University of Strasbourg IdEx program.}

\ack{We thank Adam Brandstetter-Kunc, Rodolfo A.\ Jalabert, and Dietmar Weinmann for stimulating discussions.}

\appendix
\section{Comparison between cylinders and spheroids}
\label{Sec:Comparison}

The list of analytic expressions describing the electrostatic response of NPs of different geometries is commonly thought to be limited to spheres and ellipsoids \cite{Bohren1983}. Perhaps surprisingly, and certainly unfortunately due to the experimental ubiquity of cylindrical nanorods, the plasmonic modes of cylinders cannot be found analytically via the typical route of solving Laplace's equation, finding the polarizability of the oscillating dipole, and then extracting from its pole the LSP resonance frequency. However, it was shown in the numerical study of Venermo and Sihvola \cite{Venermo2004} that the polarizability of a cylinder closely matches the closed-form expression for the polarizability of a spheroid (i.e., an ellipsoid of revolution) with the same length-to-diameter ratio.

In this appendix, we provide a comparison between two sets of analytical results for the LSP resonance frequency in the quasistatic regime: firstly, our closed-form  expressions for cylinders as derived in \S\ref{sec:1_cyl}, and secondly the well-known expressions for a spheroid that can be found, e.g., in reference \cite{Bohren1983}. 
We consider the spheroid to be described by the semi-axes $L/2$ and $a$, so that it closely resembles the cylinder considered in the main text (of length $L$ and diameter $2a$). By modulating the aspect ratio $L/a$ governing the spheroid geometry, we can describe oblate spheroids ($L < 2a$), spheres ($L = 2a$), and prolate spheroids ($L > 2a$). Crucially, spheroids share their limiting forms with cylinders: in the limit $L \ll a$ they both reduce to disks, and in the opposing limit $L \gg a$ they both become wires.

\begin{figure}[tb]
\begin{center}
 \includegraphics[width=1.0\linewidth]{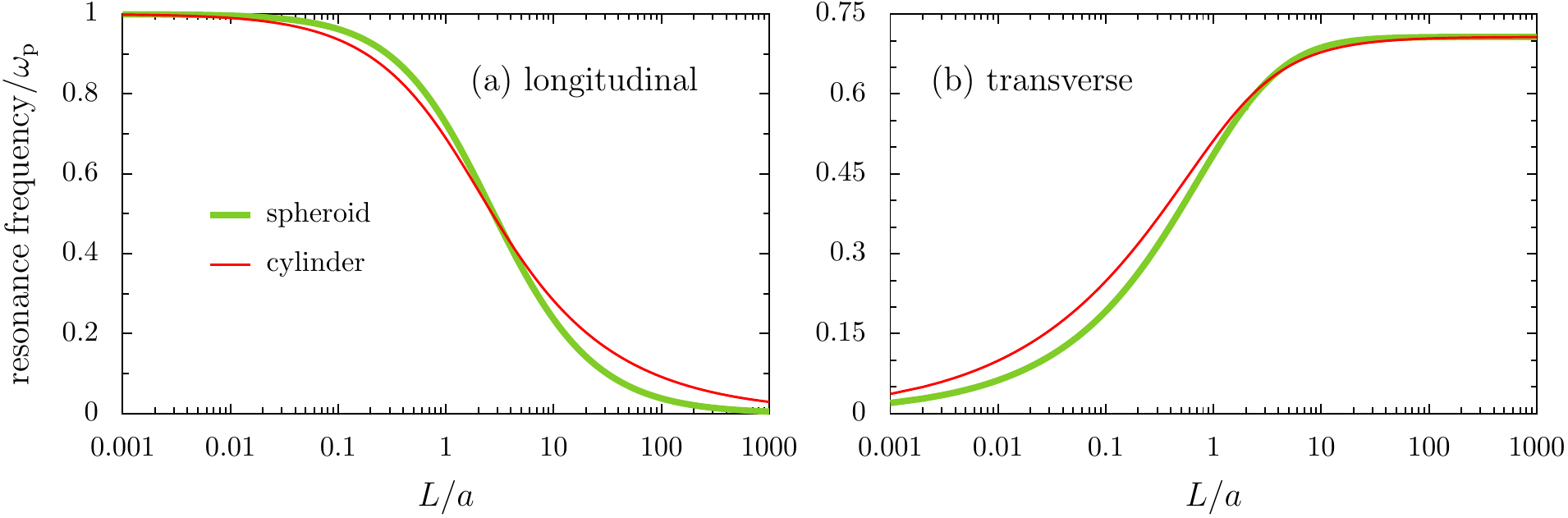}
 \caption{Resonance frequencies of the LSPs in a spheroidal (thick green lines) and cylindrical (thin red lines) nanoparticle, in units of the plasma frequency $\omega_\mathrm{p}$, as a function of the aspect ratio $L/a$. Panel (a): longitudinal mode. Panel (b): transverse mode. The analytical expressions for the cylinder are given by \eqref{eq:dimer_spectrum_1} and \eqref{eq:Trans_modes}, and for the spheroid the closed-form expressions can be found, e.g., in reference \cite{Bohren1983}.}
 \label{fig:spheroid}
 \end{center}
\end{figure}

In figure~\ref{fig:spheroid} we plot the LSP resonance frequencies as a function of the aspect ratio $L/a$, for both the longitudinal and transverse polarizations [in panels (a) and (b), respectively]. The standard textbook expressions for a spheroid are given on pages 146 and 345 of reference \cite{Bohren1983}, and the cylindrical results of this work are given by \eqref{eq:dimer_spectrum_1} and \eqref{eq:Trans_modes}. As was found in the numerical comparison of reference \cite{Venermo2004}, there is a very good agreement between the spheroid (thick green lines) and 
the cylinder (thin red lines) as they evolve geometrically. The correspondence is perfect in the extreme disk and wire limits.

The small discrepancies in resonance frequency between the two geometries can be analyzed by considering the asymptotics of the spheroid. 
In the longitudinal polarization, the resonance frequency (in units of the plasma frequency $\omega_\mathrm{p}$) approaches 
$1-\pi L/8a$ in the disk limit ($L\ll a$), and $(2 a/L) [ \ln{(L/a)} - 1]^{1/2}$ in the wire limit. Therefore, as shown in panel (a) 
of figure~\ref{fig:spheroid}, the agreement is excellent in the disk limit, since both the spheroid and cylinder frequencies scale linearly with 
$L/a$ [cf.\ \eqref{eq:disk1}]. In the wire limit ($L\gg a$), while the $L/a$ scaling is inverse linear for the spheroid, it is inverse square root for the cylinder [cf.\ \eqref{eq:wire1}]. For the transverse polarization, the spheroid asymptotics is $(\pi L/8a)^{1/2}$ in the disk limit, displaying the same square root scaling as the cylinder [cf.\ \eqref{eq:Trans_limits_disk}]. In the wire limit, the spheroid behavior is described by 
$\{1-2 (a/L)^2[\ln{(L/a)} - 1]\}/\sqrt{2}$, an inverse square behavior which differs from the inverse linear scaling for the cylinder [cf.\ \eqref{eq:Trans_limits_wire}], but does not lead to a noticeable deviation on the scale of figure~\ref{fig:spheroid}(b). 

We conclude that, as shown numerically in reference \cite{Venermo2004}, the plasmonic behavior of spheroids and cylinders is indeed very similar. Our closed-form expressions for the cylinder provide insight into the small differences between the two geometries, most prominently in the wire limit asymptotics, and challenges the long-held notion that only spheres and spheroids may be characterized analytically.

\renewcommand{\theequation}{B \arabic{equation}}
\section{Schr\"odinger equation with a cylindrical step potential}
\label{sec:app}

In this appendix, we provide details about the bound-state solutions to the Schr\"odinger equation 
\eqref{eq:Full_SE},
which enable us to evaluate semiclassically the average number of spill-out electrons \eqref{eq:N_out_def}, in both the longitudinal and transverse directions. 

Separating the variables as in \eqref{eq:psi_sep}, 
the transverse wave functions $F_{n m}(r, \theta)$ are subject to the following Schr{\"o}dinger equation, 
\begin{equation}
\label{eq:Cylinder_SE}
\left\{ \frac{\partial^2}{\partial r^2}  + \frac{1}{r}\frac{\partial}{\partial r}   + \frac{1}{r^2} \frac{\partial^2}{\partial \theta^2} 
  + \left[ k_{r }^2 - \frac{2 m_\mathrm{e}}{\hbar^2} V_r(r) \right] \right\}F_{n m}(r, \theta) = 0.
\end{equation}
With the ansatz $F_{n m}(r, \theta) = R_{n m}(r)\; \mathrm{e}^{\mathrm{i} m \theta} /(2 \pi)^{1/2}$, where the quantum number $m\in\mathbb{Z}$, and with the notation $\kappa_{r } = (k_0^2-k_{r }^2)^{1/2}$, where $k_0 = (2 m_\mathrm{e} V_0 / \hbar^2)^{1/2}$, one finds 
the following bound state solutions 
\begin{equation}
\label{eq:Cylinder_WF}
  R_{n m}(r) = C_{n m}
  \begin{cases} 
   J_{m}(k_{r } r), & r \leqslant a, \\
    \displaystyle\frac{J_{m}(k_{r } a)}{K_{m}(\kappa_{r } a)} K_{m}(\kappa_{r } r),       & r > a,
  \end{cases}
\end{equation}
where $J_{m}(x)$ and $K_{m}(x)$ are the Bessel functions of the first and second kinds, respectively. The normalization constant in \eqref{eq:Cylinder_WF} is given by 
\begin{equation}
\label{eq:Cylinder_NORM}
 C_{n m} = \frac{\sqrt{2}}{a} \left\{  \left[ \frac{J_{m}(k_{r } a)}{K_{m}(\kappa_{r } a)} \right]^2 K_{m+1}(\kappa_{r } a) K_{m-1}(\kappa_{r } a) 
 - J_{m+1}(k_{r } a) J_{m-1}(k_{r } a) \right\}^{-1/2},
\end{equation}
while the transverse motion is subject to energy quantization via the transcendental equation
$k_{r } {J_{m+1}(k_{r } a)}/{J_{m}(k_{r } a)} = \kappa_{r } {K_{m+1}(\kappa_{r } a)}/{K_{m}(\kappa_{r } a)}$, 
whose solutions are labeled with the quantum number $n$. 

The longitudinal wave functions $Z_{\tilde{n}}(z)$ entering \eqref{eq:psi_sep} obey 
\begin{equation}
\label{eq:Cart_SE}
 \frac{\mathrm{d}^2}{\mathrm{d} z^2} Z_{\tilde{n}}(z) + \left[ k_z^2 - \frac{2 m_\mathrm{e}}{\hbar^2} V_z(z) \right] Z_{\tilde{n}}(z) = 0, 
\end{equation}
which is equivalent to the textbook quantum mechanics exercise of a one-dimensional particle in a square box \cite{cohen}.
The solutions of \eqref{eq:Cart_SE} have either a symmetric ($\mathrm{s}$) or an antisymmetric ($\mathrm{a}$) parity, which we specify as $Z_{\tilde{n}}(z) = Z_{\tilde{n}, \mathrm{p}}(z)$, where the index $\mathrm{p} = (\mathrm{s, a})$. The even bound state solutions are given by
\begin{subequations}
\label{eq:Cart_WF}
\begin{equation}
\label{eq:Cart_WF_sym}
  Z_{\tilde{n}, \mathrm{s}}(z) = \sqrt{\frac{\kappa_z}{1+ {\kappa_z L}/{2}}}
  \begin{cases} 
   \displaystyle\cos \left( \frac{k_z L}{2} \right) \mathrm{e}^{\kappa_z ( {L}/{2} + z )}, & \displaystyle z \leqslant -\frac{L}{2}, \\[.3cm]
   \displaystyle \cos \left( k_z z \right), & \displaystyle |z| < \frac{L}{2}, \\[.3cm]
   \displaystyle \cos \left( \frac{k_z L}{2} \right) \mathrm{e}^{\kappa_z \left( {L}/{2} - z  \right)}, & \displaystyle z \geqslant \frac{L}{2},
  \end{cases}
\end{equation}
where $\kappa_z = (k_0^2 - k_z^2)^{1/2}>0$. Similarly, the odd solutions 
read
\begin{equation}
\label{eq:Cart_WF_anti}
  Z_{\tilde{n}, \mathrm{a}}(z) = \sqrt{\frac{\kappa_z}{1+ {\kappa_z L}/{2}}}
  \begin{cases} 
    \displaystyle-\sin \left( \frac{k_z L}{2} \right) \mathrm{e}^{\kappa_z ( {L}/{2} + z  )}, &  \displaystyle z \leqslant -\frac{L}{2}, \\[.3cm]
     \displaystyle \sin \left( k_z z \right), & \displaystyle |z| < \frac{L}{2}, \\[.3cm]
     \displaystyle \sin \left( \frac{k_z L}{2} \right) \mathrm{e}^{\kappa_z ( {L}/{2} - z  )}, &  \displaystyle z \geqslant \frac{L}{2}.
  \end{cases}
\end{equation}
\end{subequations}
Both sets of eigenfunctions~\eqref{eq:Cart_WF_sym} and~\eqref{eq:Cart_WF_anti} are associated with an individual transcendental equation describing the quantization of energy due to the longitudinal confinement, explicitly  
$\tan{( {k_z L}/{2})} = {\kappa_z}/{k_z}$ ({$\mathrm{s}$ modes}) 
and
$\tan{( {k_z L}/{2})} = -{k_z}/{\kappa_z}$ ($\mathrm{a}$ modes).
The solutions of these equations are labeled with the third quantum number of the problem, $\tilde{n}$.

Now that the full Schr{\"o}dinger equation \eqref{eq:Full_SE} is solved, we proceed with the evaluation of 
the integrals entering \eqref{eq:N_out_def}, namely
\begin{equation}
\label{eq:electron_integrals1}
 \mathcal{R}_{n m}^{\mathrm{in}} = \int_0^a  \mathrm{d}{r}\; r |R_{n m} (r)|^2 , \qquad\qquad
 \mathcal{R}_{n m}^{\mathrm{out}} = \int_a^{\infty} \mathrm{d}{r}\; r |R_{n m} (r)|^2, 
 \end{equation}
 \begin{equation}
 \label{eq:electron_integrals3}
 \mathcal{Z}_{\tilde{n}, \mathrm{p}}^{\mathrm{in}} = \int_{-L/2}^{+L/2}\mathrm{d}{z}\; |Z_{\tilde{n}, \mathrm{p}} (z)|^2 ,  \qquad\qquad
 \mathcal{Z}_{\tilde{n}, \mathrm{p}}^{\mathrm{out}} = \left( \int_{-\infty}^{-L/2} + \int_{+L/2}^{+\infty} \right)\mathrm{d}{z}\; |Z_{\tilde{n}, \mathrm{p}} (z)|^2 , 
 \end{equation}
which describe the probability of finding the electrons inside or outside the cylindrical NP, in either the transverse or longitudinal directions. 
With \eqref{eq:Cylinder_WF}, we obtain for the transverse integrals \eqref{eq:electron_integrals1} the results
\begin{subequations}
\label{eq:electron_integrals_2}
\begin{align}
 &\mathcal{R}_{n m}^{\mathrm{in}} = \frac{J_{m+1}(k_{r } a) J_{m-1}(k_{r } a) - J_{m}^2(k_{r } a)}{
   J_{m+1}(k_{r } a) J_{m-1}(k_{r } a) - J_{m}^2(k_{r } a) 
   K_{m+1} (\kappa_{r } a) K_{m-1} (\kappa_{r } a)/K_{m}^2 (\kappa_{r } a)}
   , \\
& \mathcal{R}_{n m}^{\mathrm{out}} = \frac{K_{m+1}(\kappa_{r } a) K_{m-1}(\kappa_{r } a) - K_{m}^2(\kappa_{r } a)}{K_{m+1}(\kappa_{r } a) K_{m-1}(\kappa_{r } a) - K_{m}^2(\kappa_{r } a) 
J_{m+1} (k_{r } a) J_{m-1} (k_{r } a)/J_{m}^2 (k_{r } a)
}.
 \end{align}
\end{subequations}
Similarly, using \eqref{eq:Cart_WF}, we find for the longitudinal integrals \eqref{eq:electron_integrals3} 
\begin{equation}
\label{eq:electron_integrals_2_b}
 \mathcal{Z}_{\tilde{n}, \mathrm{p}}^{\mathrm{in}} = \frac{1}{1+{\kappa_z L}/{2}} \left( \frac{\kappa_z L}{2} + \frac{\kappa_z^2}{k_0^2} \right), \qquad\qquad
  \mathcal{Z}_{\tilde{n}, \mathrm{p}}^{\mathrm{out}} = \frac{k_z^2/k_0^2}{1+{\kappa_z L}/{2}}. 
 \end{equation}

In the high-energy semiclassical limit ($k_0 a \gg 1, k_0 L \gg 1$), which is well suited for the problem at hand \cite{Weick2006}, we find that the expressions \eqref{eq:electron_integrals_2} and \eqref{eq:electron_integrals_2_b} are well-approximated by
\begin{equation}
\label{eq:electron_integrals_3}
 \mathcal{R}_{n m}^{\mathrm{in}} \simeq 1, \qquad\qquad
 \mathcal{R}_{n m}^{\mathrm{out}} \simeq \frac{1}{2} \frac{1}{\kappa_{r } a}, \qquad\qquad
 \mathcal{Z}_{\tilde{n}, \mathrm{p}}^{\mathrm{in}} \simeq 1, \qquad\qquad
 \mathcal{Z}_{\tilde{n}, \mathrm{p}}^{\mathrm{out}} \simeq \frac{2}{\kappa_{z} L} \frac{k_z^2}{k_0^2}, 
 \end{equation}
which then lead to \eqref{eq:electronic_int_approx}.\footnote{In this semiclassical limit, the leading order expressions \eqref{eq:electron_integrals_3} do not satisfy unitarity, which requires the inclusion of high-order terms. However, since the absent terms are of negligible importance for the range of parameters we consider in this work we may omit them. Notably, this semiclassical limit has been shown to be an excellent approximation for a spherical NP \cite{Weick2006} and it has the significant advantage of providing additional physical insight.}

\renewcommand{\theequation}{C \arabic{equation}}
\section{Toy model: two coupled oscillating dipoles}
\label{sec:dipole}
In this appendix, we demonstrate that the results \eqref{eq:dimer_spectrum} and \eqref{eq:dimer_spectrum_trans} for the resonance frequencies of 
the coupled modes in a dimer of cylindrical NPs in the limit of large interparticle separation distance 
(i.e., $d/a\gg1$ and $d/L\gg1$) can be recovered from a toy model of two coupled anisotropic oscillating dipolar moments. 

Let us consider two ideal electric dipoles $\mathbf{p}_i=-N_\mathrm{e}e\,\mathbf{r}_i$ ($i=1,2$), 
with $\mathbf{r}_i$ the associated displacement of the electronic cloud with charge $-N_\mathrm{e}e$ and 
mass $N_\mathrm{e}m_\mathrm{e}$. The dimer (with interparticle distance $d$) is aligned along the $z$ direction and each dipole oscillates at the frequency $\omega_{0,\parallel}^\mathrm{dip}$ in the longitudinal ($z$) direction and $\omega_{0,\perp}^\mathrm{dip}$ in the transverse ($x,y$) directions. The Lagrangian for the system described above reads
\begin{equation}
\label{eq:lagrangian}
\mathcal{L}=\frac{N_\mathrm{e}m_\mathrm{e}}{2}
\sum_{i=1}^2
\left[\dot{\mathbf{r}}_i^2
-{\omega_{0,\perp}^\mathrm{dip}}^2\left(x_i^2+y_i^2\right)
-{\omega_{0,\parallel}^\mathrm{dip}}^2z_i^2\right]
-\frac{N_\mathrm{e}^2e^2}{d^3}\left[\mathbf{r}_1\cdot\mathbf{r}_2
-3\left(\mathbf{r}_1\cdot\hat{z}\right)\left(\mathbf{r}_2\cdot\hat{z}\right)\right].
\end{equation} 
Using that $N_\mathrm{e}e^2/m_\mathrm{e}=\omega_\mathrm{p}^2V/4\pi$, with $V$ the volume of the electronic cloud, the Euler--Lagrange equations of motion for the toy model \eqref{eq:lagrangian} leads to
the coupled mode resonance frequencies
\begin{equation}
\omega_{\tau,\parallel}^\mathrm{dip}=\sqrt{{\omega_{0,\parallel}^\mathrm{dip}}^2+2\tau\omega_\mathrm{p}^2\frac{V}{4\pi d^3}}, 
\qquad
\omega_{\tau,\perp}^\mathrm{dip}=\sqrt{{\omega_{0,\perp}^\mathrm{dip}}^2+\tau\omega_\mathrm{p}^2\frac{V}{4\pi d^3}}, \qquad \tau=\pm.
\end{equation}
With $V=\pi a^2L$ the volume of the cylinder considered in the main text, 
the expressions above correspond to \eqref{eq:dimer_spectrum} and \eqref{eq:dimer_spectrum_trans} with $\Omega$ given by 
\eqref{eq:Omega_approx} in the limit of $d/a\gg1$ and $d/L\gg1$.


\end{document}